\documentclass[twocolumn,prd,showpacs]{revtex4}

\usepackage{amsmath}

\begin{document}

\title{New positive small vacuum region
gravitational energy expressions }

\author{Lau Loi So$^{1,4}$}
\email{s0242010@webmail.tku.edu.tw}
\author{James M. Nester$^{1,2,3}$}
\email{nester@phy.ncu.edu.tw}

 \affiliation{$^1$Department of Physics,
 National Central University, Chungli 320, Taiwan\\
 $^2$Graduate Institute of Astronomy,
 National Central University, Chungli 320, Taiwan\\
 $^3$Center for Mathematics and Theoretical Physics,
National Central University, Chungli 320, Taiwan\\
$^4$Current address: Department of Physics, Tamkang University,
Tamsui 251, Taiwan}


\begin{abstract}
\noindent We  construct an infinite number of new holonomic
quasi-local gravitational energy-momentum density pseudotensors with
good limits asymptotically and in small regions, both materially and
in vacuum. For small vacuum regions they are all a positive multiple
of the Bel-Robinson tensor and consequently have positive energy.
\end{abstract}

\maketitle


The localization of energy for gravitating systems has been an
important fundamental problem since before Einstein finalized his
field equations (see \cite{JR06}). In recent decades there has been
some significant progress on this outstanding problem. New
perspectives and ideas on gravitational energy continue to be
considered as noteworthy (e.g., \cite{SY79,CNC99,AGP00,LY03}).
Recall that it is a well-known inevitable consequence of the
equivalence principle that there is
 no covariant (reference frame independent) description for the
gravitational energy-momentum density (for a discussion see Ch.\ 20
in \cite{MTW}).  The coordinate dependent pseudotensor density
description used in earlier times has in recent decades largely been
replaced by the quasi-local perspective (i.e., energy-momentum is to
be associated with a closed 2-surface; for a comprehensive review
see \cite{Sza04}). However it has been noted that the Hamiltonian
approach to quasi-local energy-momentum includes all the possible
pseudotensors simply by taking their associated superpotentials as
the Hamiltonian boundary term. This not only shows the pseudotensors
to be a special type of quasi-local expression but, moreover,
reveals their ambiguities to be just the same as those of the
quasi-local Hamiltonian boundary term. The Hamiltonian approach,
however, gives these ambiguities a clear physical and geometric
meaning. Specifically, the freedom in the choice of expression is
associated (via the boundary term in the Hamiltonian variation) with
the freedom to choose the type of boundary conditions, and the
choice of quasi-local reference (which determines the zero energy or
ground state) is effectively just the choice of coordinates on the
boundary for the holonomic pseudotensor
\cite{CN99,CN00,beij,Nes04,CNT05}.

Various criteria have been proposed for quasi-local quantities (see,
e.g., \cite{Sza04,LY06}), including having the correct asymptotic
limits at infinity and positivity. Positivity has been regarded as a
strong condition. It is difficult to prove the positivity of the
quasi-local energy determined by some expression for a general
region. Just considering a small vacuum region already gives a
significant requirement, namely that the energy-momentum be
proportional to the Bel-Robinson tensor \cite{DFS99,Sza04}.  This
will guarantee positive energy within a small vacuum region.

 We have had some hope that positivity within a small vacuum region would be a quite strong restriction,
perhaps even selecting a unique expression. We found that it is
indeed a serious constraint; in particular the requirement of a
positive energy density eliminates {\em all} of the classical
pseudotensors, nevertheless we found that it still allows a
surprising amount of freedom. Specifically, we recently
\cite{icga7so,So06,SN08a,SN08b} showed that none of the classical
pseudotensors (Einstein \cite{Tra62}, Papapetrou
\cite{Papa48,Gupta,Jackiw}, Landau-Lifshitz \cite{LL},
Bergman-Thompson \cite{BT53}, M{\o}ller \cite{Mol58}, Goldberg
\cite{Gol58}, Weinberg \cite{Wein72}) yields a positive energy
density within a small vacuum region, although certain linear
combinations of them do.

Here we show how to construct an infinite number (more precisely an
11 parameter class) of new quasi-local pseudotensorial expressions
which, in addition to having the proper asymptotic and small
material limit, have a small region energy-momentum density
proportional to the Bel-Robinson tensor.  This guarantees a locally
positive energy density for small vacuum regions. We will present
our new results after reviewing some technical background.

{\em Riemann normal coordinates (RNC).\/} As Riemann first argued
(see, e.g., \cite{Spivak}), at any preselected point one can choose
coordinates such that at the point, (i) $x^\mu=0$, (ii) the metric
coefficients have the standard flat values, (iii) the first
derivatives of the metric vanish, (iv) and the second derivatives
have the minimum number (20 for $n=4$) of independent values.
Specifically $g_{\alpha\beta}|_0=\bar g_{\alpha\beta}$,
$\partial_\mu g_{\alpha\beta}|_0=0$,
$3\partial_{\mu\nu}g_{\alpha\beta}|_0=-(R_{\alpha\mu\beta\nu}+R_{\alpha\nu\beta\mu})|_0$,
where $R^\alpha{}_{\beta\mu\nu}$ is the Riemannian curvature tensor
(all indicies are holonomic, our conventions, unless otherwise
specified, follow MTW \cite{MTW}), and, in our case, $\bar
g_{\alpha\beta}=\eta_{\alpha\beta}=\hbox{diag}(-1,+1,+1,+1)$ is the
Minkowsi spacetime metric.  The corresponding Levi-Civita connection
values are $ \Gamma^\alpha{}_{\beta\gamma}|_0=0$,  $3\partial_\mu
\Gamma^\alpha{}_{\beta\nu}|_0=-(R^\alpha{}_{\beta\nu\mu}+R^\alpha{}_{\nu\beta\mu})|_0$.

{\em Quadratic curvature basis.\/} It turns out that when expanded
in RNC the lowest non-vanishing vacuum energy-momentum expressions
are of the second order and are quadratic in the curvature tensor:
$t_{\mu\nu}\sim(R_{\cdot\cdot\cdot\cdot}R^\cdot{}_\cdot{}^\cdot{}_\cdot)_{\mu\nu
ij}x^ix^j$. An investigation \cite{DFS99} of all such possible terms
(taking into account the Weyl = vacuum Riemann tensor symmetries)
shows that they can be written in terms of
$Q_{\mu\alpha\nu\beta}:=R_{a\mu b\alpha}R^a{}_\nu{}^b{}_\beta\equiv
Q_{\nu\beta\mu\alpha}\equiv Q_{\alpha\mu\beta\nu}$, where all the
symmetries have been indicated. The cited work defined the three
basis combinations $
X_{\mu\nu\alpha\beta}:=2Q_{\alpha(\mu\nu)\beta}$,
$Y_{\mu\nu\alpha\beta}:=2Q_{\alpha\beta(\mu\nu)}$,
$Z_{\mu\nu\alpha\beta}:=Q_{\alpha\mu\beta\nu} +
Q_{\alpha\nu\beta\mu}$, along with the trace tensor
$T_{\mu\nu\alpha\beta}:=-\frac16
g_{\mu\nu}Q^\sigma{}_{\alpha\beta\sigma}$. One could write all our
expansions as linear combinations of $X$, $Y$, $Z$, $T$.  However it
is more suitable for physical purposes to use the Bel-Robinson
tensor.

{\em The Bel-Robinson and some related tensors.\/} The Bel-Robinson
tensor has many well known remarkable properties, see e.g.,
\cite{DFS99,Gar01}.  For our considerations we are interested in it
only in the vacuum, where the Riemann tensor reduces to the Weyl
tensor. In this case the Bel-Robinson tensor is completely symmetric
and traceless. In vacuum the Bel-Robinson tensor $B$ and two other
convenient tensors $S$ and $K$
 are given
in terms of the aforementioned basis for quadratic terms by
\begin{equation}
B:=Z+3T, \quad S:=-2X+2Z-6T, \quad K:=Y+9T.
\end{equation}
(Here and below we often suppress obvious indicies.) Our leading
order non-vanishing vacuum expressions could be given as linear
combinations of $B$, $S$, and $K$. As can be directly verified,
these three combinations satisfy the second order divergence free
condition
\begin{equation}
\partial_\beta (x^i x^j t_{ij}{}^{\alpha\beta})\equiv2x^jt_{\beta j}{}^{\alpha\beta}\equiv0,
\end{equation}
and all such tensors are some linear combination of these three.
While the tensor $S$ has been known for a long time, the tensor $K$
(which was recently introduced \cite{So06,icga7so,SN08a}) also
enjoys this divergence free property.  However, for our purposes
here we have found it even more convenient to use instead of $K$ the
tensor
\begin{equation}
V:=-2X+Y+2Z+3T\equiv S+K,
\end{equation}
which, in addition to enjoying the vanishing divergence property, as
we shall see also shares an important positive energy property with
the Bel-Robinson tensor \cite{SN08b}.

{\em All possible superpotentials}. One can always express a
conserved energy-momentum expression in terms of some suitable {\em
superpotential} in the form
\cite{Mol58,Gol58,Tra62,Nes04,CNC99,SN08a}
\begin{equation}
t^\mu{}_\nu:=\partial_\lambda U^{[\mu\lambda]}{}_\nu;
\end{equation} the anti-symmetry guarantees the conserved current condition:
$\partial_\mu t^\mu{}_\nu=0$. Properly the superpotential should be
a density of weight one and the energy-momentum density should be a
mixed tensor density of the indicated type. The physical constraints
are such that in the weak field, asymptotic, and material limits we
want $U$ to be {\it linear} in $\Gamma$ or, equivalently, $
\partial g$. We consider only superpotentials with this property.
Using the metric and connection one can form only three linearly
independent terms:
\begin{equation}
|g|^{1\over2}g^{\sigma[\mu}\Gamma^{\lambda]}{}_{\sigma\nu}, \
|g|^{1\over2}\delta^{[\mu}_\nu\Gamma^{\lambda]}{}_{\alpha\sigma}g^{\alpha\sigma},
\ |g|^{1\over2}\delta^{[\mu}_\nu
g^{\lambda]\gamma}\Gamma^\alpha{}_{\alpha\gamma}.
\end{equation}
One of these simple choices,
$2|g|^{1\over2}g^{\beta[\mu}\Gamma^{\lambda]}{}_{\beta\nu}$, is the
well-known M{\o}ller \cite{Mol58} superpotential, which has an
incorrect small region matter interior limit \cite{SN08a}.

Consider linear combinations of the three terms.  In the small
region matter interior limit the derivative of each term gives rise
in zeroth order to a certain linear combination of
$|g|^{\frac12}R^\mu{}_\nu$ and $|g|^{\frac12}R\delta^\mu{}_\nu$. The
desired limit to this order is $2|g|^{\frac12}G^\mu{}_\nu$. (Using
Einstein's equation this will become $2\kappa
|g|^{\frac12}T^\mu{}_\nu$, the limit required by the {\em
equivalence principle}.) This requirement imposes two constraints on
the three parameters, hence there is a one parameter set which is
satisfactory in this limit, namely
\begin{equation}
2|g|^{1\over2}\left[\Gamma^{[\mu\lambda]}{}_\nu +(3-2k)
\Gamma^{\gamma[\mu}{}_\gamma\delta^{\lambda]}_\nu+
k\delta^{[\mu}_\nu\Gamma^{\lambda]\gamma}{}_\gamma\right].
\end{equation}
 However
consideration of obtaining the proper energy-momentum asymptotic
limit (described in  \S 20.2 in \cite{MTW}) removes this freedom,
giving (uniquely up to terms which vanish in this limit) a familiar
expression, Freud's superpotential \cite{Freud} for the Einstein
pseudotensor:
\begin{eqnarray}
U_{\rm
FE}^{\mu\lambda}{}_\nu&:=&|g|^{1\over2}g^{\beta\sigma}\Gamma^\alpha{}_{\beta\gamma}\delta^{\mu\lambda\gamma}_{\alpha\sigma\nu}\nonumber\\
&=& 2|g|^{1\over2}\left(\Gamma^{[\mu\lambda]}{}_\nu +
\Gamma^{\gamma[\mu}{}_\gamma\delta^{\lambda]}_\nu+
\delta^{[\mu}_\nu\Gamma^{\lambda]\gamma}{}_\gamma\right),\label{Freud}
\end{eqnarray}
 which has good limits both at infinity and for small regions
inside of matter.

Using only the metric and its first derivatives, of the classic
pseudotensors, one can construct only the Freud-Einstein and
M{\o}ller superpotential expressions. However it should be noted
that these expressions really only make sense in nearly Cartesian
coordinates, hence there actually is an implicit background
Minkowski structure, a reference geometry $\bar g, \bar\Gamma$.  If
we explicitly introduce this structure we can go much further. In
this work we will use coordinates which are asymptotically flat
Minkowski for large and small distances. From now on indicies will
be transvected only with the flat metric ${\bar g}$.

A general superpotential can be expressed as an expansion in terms
of powers of $h:=g-\bar g$ in the form $U=U_0+U_1(h)+U_2(h^2)
+\cdots$ where $U_k$ is linear in $\Gamma\sim\partial g$.  It is
easy to see that for the limits of interest we need not consider the
terms for $k\ge2$.  It is well known  that the Freud superpotential
(\ref{Freud})  gives good values to linear order. It has the correct
limit both at infinity and within matter to zeroth order. Without
loss of generality we can take $U_0$ to be Freud (alternate choices
would just shift our expression for $U_1$). Now we want to consider
expressions of the form
\begin{equation}
U_1^{[\mu\lambda]}{}_\nu=F_\nu{}^{[\lambda\mu](ij)(bc)}{}_ah_{ij}\Gamma^a
{}_{bc},
\end{equation}
where $F$ is some constant tensor made from ${\bar g}_{\alpha\beta}$
and its inverse with the indicated symmetries. We give all possible
terms to this order. There are 7 terms which include traces:
$
h_\nu{}^{[\mu}\Gamma_a{}^{\lambda]a}$, 
$h_\nu{}^{[\mu}\Gamma^{\lambda]a}{}_a$, 
$h^a{}_a\Gamma^{[\mu\lambda]}{}_\nu$, 
$h^a{}_a\delta^{[\mu}_\nu\Gamma_a{}^{\lambda]a}$, 
$\delta^{[\mu}_\nu h^{\lambda]c}\Gamma^a{}_{ac}$, 
$\delta^{[\mu}_\nu h^{\lambda]}{}_a\Gamma^{ab}{}_b$, 
$h^a{}_a\delta^{[\mu}_\nu \Gamma^{\lambda]b}{}_b$, 
consequently at second order they give rise to terms proportional to
the Ricci tensor, and hence they vanish in vacuum.  They need not be
given any more detailed analysis here. However there are 6 more
terms,
\begin{eqnarray}
U_{\textsf{A}}:=2h^{a[\mu}\Gamma_a{}^{\lambda]}{}_\nu,&\qquad&
U_{\textsf{B}}:=2h^{b[\mu}\Gamma^{\lambda]}{}_{b\nu},\nonumber\\
U_{\textsf{C}}:=2h^{b[\mu}\Gamma_\nu {}^{\lambda]}{}_b,&\qquad&
U_{\textsf{D}}:=2h^{ba}\Gamma_a {}_b{}^{[\mu}\delta^{\lambda]}_\nu,\nonumber\\
U_{\textsf{E}}:=2h^{bc}\delta^{[\mu}_\nu\Gamma^{\lambda]}{}_{bc},&\qquad&
U_{\textsf{F}}:=2h_{b\nu}\Gamma^{[\mu}{}^{\lambda]b},\label{UAF}
\end{eqnarray}
which give rise to the indicated non-vanishing vacuum contributions
to the energy-momentum density at the second order:
\begin{eqnarray}
 \partial U_A&\simeq&{\textstyle{x^ix^j\over9\cdot2}}[-B+\frac{1}{2}S+2V]_{ij}{}^\mu{}_\nu,\label{tAA}\\
 \partial U_B&\simeq&{\textstyle{x^ix^j\over9\cdot2}}[-B-S-V]_{ij}{}^\mu{}_\nu,\label{tBB}\\
 \partial U_C&\simeq&{\textstyle{x^ix^j\over9\cdot2}}[2B+{1\over2}S-V]_{ij}{}^\mu{}_\nu,\label{tCC}\\
 \partial U_D&\simeq&{\textstyle{x^ix^j\over9\cdot2}}[2B-{3\over2}S+V]_{ij}{}^\mu{}_\nu,\label{tDD} \\
 \partial U_E&\simeq&{\textstyle{x^ix^j\over9\cdot2}}[4B-3S+2V]_{ij}{}^\mu{}_\nu,\label{tEE}\\
\partial
U_F&\simeq&{\textstyle{x^ix^j\over9\cdot2}}[3B+{3\over2}S]_{ij}{}^\mu{}_\nu.\label{tFF}
\end{eqnarray}
(It turns out that two of the expressions produce proportional
results to this order.)

 Any linear combinations of the expressions (\ref{UAF}),
 $aU_{\textsf{A}}+bU_{\textsf{B}}+cU_{\textsf{C}}+dU_{\textsf{D}}+eU_{\textsf{E}}+fU_{\textsf{F}}$,
 may be added to the Freud expression.  The Freud expression (normalized to give the correct asymptotic and small region
non-vacuum values) leads to a small region vacuum second order
contribution of \cite{MTW,Gar73,DFS99}
\begin{equation}
\partial_\lambda U_{\rm FE}^{[\mu\lambda]}{}_\nu\simeq
{\textstyle{x^ix^j\over9\cdot2}}[4B-S]_{ij}{}^\mu{}_\nu.
\end{equation}

The resultant total energy-momentum expression in vacuum to second
order then has the form
\begin{equation}
\partial_\lambda U^{\mu\lambda}{}_\nu\simeq {\textstyle{x^ix^j\over9\cdot2}}[\beta B+sS
+vV]_{ij}{}^\mu{}_\nu,
\end{equation}
where
\begin{eqnarray}
\beta&=&4-a-b+2c+2(d+2e) +3f,\\
s&=&-1+\frac12a-b+\frac12c-\frac32(d+2e)+\frac32f,\\
v&=&2a-b-c+(d+2e).
\end{eqnarray}

 The usual
 requirement is that one should have a positive multiple of the
Bel-Robinson tensor \cite{DFS99}, which means that the parameter
combinations $s$ and $v$ should vanish, and  $\beta$, the
coefficient of $B$, should be positive.
  This means, in addition to one
inequality, 2 restrictions on the 6 parameters controlling the
contributions from (\ref{UAF}).  Thus there are an infinite number
of expressions, controlled by 11 parameters (4 free parameters
associated with these expressions plus 7 additional parameters
associated with the expressions that explicitly included traces),
which have  {\em positive} energy to this order.

Explicitly we can eliminate and take say $a,b,d,e$ as parameters. We
then find the two parameter constraints $3f=2+3(b-a)+2(d+2e)$, $
c=2a-b+(d+2e)$ and the inequality needed for positivity is
$d+2e\ge-1$. One simple choice, $a=b=c=d=e=0$ and $f=2/3$, yields
the one case which had been identified some time ago \cite{DFS99},
normalized here to give both the proper small region matter interior
value as well as the correct asymptotic limit.  This choice results
in a small region vacuum energy-momentum density of
\begin{equation}
2\kappa
t^{\mu\nu}={2\over3}B^{\mu\nu}{}_{ij}{x^ix^j\over2}+{\cal{O}}(x^3).
\end{equation}

We have a slightly more general result which can conveniently be
expressed in another way. Our parameterized good superpotential can
be put in the form
\begin{eqnarray}
U&=&[U_{\rm FE}-(U_{\textsf{C}}+U_{\textsf{D}})]
+\beta\left[{3\over2}(U_{\textsf{C}}+U_{\textsf{D}})+U_{\textsf{F}}\right]\quad\nonumber\\&&+v\left[2(U_{\textsf{A}}-
U_{\textsf{B}})-\frac32(U_{\textsf{C}}-U_{\textsf{D}})\right]+e[U_{\textsf{E}}-2U_{\textsf{D}}]\nonumber\\
&&+a[U_{\textsf{A}}+2U_{\textsf{C}}-U_{\textsf{F}}]+b[U_{\textsf{B}}-U_{\textsf{C}}+U_{\textsf{F}}].
\end{eqnarray}
 Here the first bracket gives the correct material and asymptotic
limit while its 2nd order vacuum contribution vanishes; the second
bracket gives a vanishing asymptotic and material limit and the
desired $Bxx/2$ in vacuum; the third bracket gives $Vxx/2$ in
vacuum, we retain this part but have required the $Sxx/2$
contribution to vanish.
 The remaining 3 terms vanish to these orders, as do the
seven terms which one could also include built from the
aforementioned explicit trace expressions.

The associated energy-momentum density for a small region, accurate
to zeroth order in matter and to second order in vacuum, is
\begin{equation}
2\kappa{\cal T}^\mu{}_\nu\simeq
2\kappa|g|^{\frac12}T^\mu{}_\nu+\beta\frac12 B_{ij}{}^\mu{}_\nu
x^ix^j+v\frac12 V_{ij}{}^\mu{}_\nu x^ix^j.
\end{equation}

The energy-momentum within a small coordinate sphere in vacuum is
\begin{equation}
P_\mu=(-E, \vec P)=\int {\cal T}^0{}_\mu d^3x.
\end{equation}
Using
\begin{equation}
\int x^a x^bd^3x =\frac13 \delta^{ab}\int
r^2d^3x=\frac{4\pi}{3\cdot5}\delta^{ab}r^5,
\end{equation}
with $a,b,c=1,2,3$, we find
\begin{eqnarray}
P^\mu&=&\frac{4\pi}{2\cdot3\cdot5}(\beta B+v
V)_a{}^{a0}{}_\nu \label{B+V}\\
&=&(\beta+v)\frac{4\pi}{30}(E^{ab}E_{ab}+H^{ab}H_{ab},
2\epsilon^{cba}E_{bd}H_a{}^d),\nonumber
\end{eqnarray}
where we have used the remarkable recently discovered identity
\cite{SN08b}, $V_a{}^{a0}{}_\nu\equiv B_a{}^{a0}{}_\nu$, along with
the traceless property,
$0=B_\mu{}^{\mu0}{}_\nu=B_0{}^{00}{}_\nu+B_a{}^{a0}{}_\nu$, and the
well-known values of certain components of the Bel-Robinson tensor
expressed in terms of the {\em electric} and {\em magnetic}
components of the Weyl tensor. From this result it is apparent that
we are guaranteed to have positive energy (in fact a non-spacelike
energy-momentum 4-vector) in the small sphere limit to second order
as long as $\beta+v>0$.

 Going to higher orders
one can expect to find a similar situation. There will be a few
positivity constraints on the terms in the power series expansion,
but there will also be many additional parameters controlling the
amplitudes of the numerous possible higher order terms in the
superpotential. Indeed we can readily see how to construct 13 second
order in $h_{\mu\nu}$ terms just by multiplying the first order ones
by $h^\alpha{}_\alpha$. Moreover we can in these 13 for any
occurrence of $h_{\mu\nu}$ make the replacement $h_{\mu\nu}\to
h_{\mu\gamma}h^\gamma{}_\nu$. That will generate many more terms.
Looking to higher orders, we can make similar adjustments in
$U_kh^k$ to get many $U_{k+1}h^{k+1}$ terms. It seems that one could
continue the process without limit. Going one or two more steps
further, which we have not yet done, might be helpful for
identifying a closed form expression that would be positive to all
orders. One might, after considerable effort, in this way find many
expressions that had positive quasi-local energy for a region of any
size.

While we have found an infinite number of new quasi-local holonomic
pseudotensor energy-momentum expressions with a certain desirable
property, yet, as far as we have investigated them, none of these
holonomic expressions appears to be natural or especially appealing.
We can offer an alternative: the teleparallel gauge current (see,
e.g., \cite{AGP00}) is {\em naturally} proportional to the
Bel-Robinson tensor in the small vacuum region limit \cite{SNC09}.

It should be remarked that our analysis also applies to many
quasi-local gravitational energy expressions which, while not
formulated within the pseudotensor framework, nevertheless take the
same form in these limits.   We have shown that, while the small
region quasi-local energy positivity Bel-Robinson requirement is a
strong condition (in particular, it excludes all the classical
pseudotensors), nevertheless it still allows for a lot of freedom in
the choice of quasi-local energy-momentum expression.

We thank the National Science Council of the R.O.C. (Taiwan) for
their support under the grants NSC 94-2112-M008-038,
95-2119-M008-027, 96-2112-M-008-005, and 97-2112-M-008-001.   JMN
was also supported in part by the National Center of Theoretical
Sciences.

\end{document}